\documentclass[sn-basic]{sn-jnl}% Basic Springer Nature Reference Style/Chemistry Reference Style
\usepackage{graphicx}%
\usepackage{multirow}%
\usepackage{amsmath,amssymb,amsfonts}%
\usepackage{amsthm}%
\usepackage{mathrsfs}%
\usepackage[title]{appendix}%
\usepackage{textcomp}%
\usepackage{manyfoot}%
\usepackage{booktabs}%
\usepackage{algorithm}%
\usepackage{algorithmicx}%
\usepackage{algpseudocode}%
\usepackage{listings}%
\usepackage{colortbl}
\usepackage[table]{xcolor}
\definecolor{blueP}{cmyk}{0.4,0.1,0,0.4}
\definecolor{maroon}{cmyk}{0,0.87,0.68,0.32}

\theoremstyle{thmstyleone}%
\theoremstyle{thmstyletwo}%

\theoremstyle{thmstylethree}%

\begin{document}

\title[Article Title]{Augmented reality for upper limb rehabilitation: real-time kinematic feedback with HoloLens 2}


%% PROPOSTA ORDINE
\author*[1,2,3]{\fnm{Beatrice} \sur{Luciani}}\email{beatrice.luciani@polimi.it}

\author[1,2]{\fnm{Alessandra} \sur{Pedrocchi}}\email{alessandra.pedrocchi@polimi.it}

\author[4]{\fnm{Peppino} \sur{Tropea}}\email{p.tropea@casadicuraigea.it}

\author[4]{\fnm{Agnese} \sur{Seregni}}\email{a.seregni@casadicuraigea.it}

\author[1,3]{\fnm{Francesco} \sur{Braghin}}\email{francesco.braghin@polimi.it}

\author[1,3]{\fnm{Marta} \sur{Gandolla}}\email{marta.gandolla@polimi.it}

\affil[1]{\orgdiv{WE-COBOT laboratory, Polo Territoriale di Lecco}, \orgname{Politecnico di Milano}, \orgaddress{\street{ Via G.Previati 1/c}, \city{Lecco}, \postcode{23900}, \country{Italy}}}

\affil[2]{\orgdiv{NEARLab - NeuroEngineering and Medical Robotics Laboratory}, \orgname{Politecnico di Milano}, \orgaddress{\street{ piazza Leonardo da Vinci 32}, \city{Milan}, \postcode{20133}, \country{Italy}}}

\affil[3]{\orgdiv{Department of Mechanical Engineering}, \orgname{Politecnico di Milano}, \orgaddress{\street{Via La Masa 1}, \city{Milan}, \postcode{20156}, \country{Italy}}}

\affil[4]{\orgdiv{Department of Neurorehabilitation Sciences}, \orgname{Casa Di Cura Igea}, \orgaddress{\street{Via Dezza 48}, \city{Milan}, \postcode{20144}, \country{Italy}}}


\abstract{Exoskeletons for rehabilitation can help enhance motor recovery in individuals suffering from neurological disorders. Precision in movement execution, especially in arm rehabilitation, is crucial to prevent maladaptive plasticity. However, current exoskeletons, while providing arm support, often lack the necessary 3D feedback capabilities to show how well rehabilitation exercises are being performed. This reduces therapist acceptance and patients' performance. Augmented Reality technologies offer promising solutions for feedback and gaming systems in rehabilitation. In this work, we leverage HoloLens 2 with its advanced hand-tracking system to develop an application for personalized rehabilitation. Our application generates custom holographic trajectories based on existing databases or therapists' demonstrations, represented as 3D tunnels. Such trajectories can be superimposed on the real training environment. They serve as a guide to the users and, thanks to colour-coded real-time feedback, indicate their performance. To assess the efficacy of the application in improving kinematic precision, we tested it with 15 healthy subjects. Comparing user tracking capabilities with and without the use of our feedback system in executing 4 different exercises, we observed significant differences, demonstrating that our application leads to improved kinematic performance. 
12 clinicians tested our system and positively evaluated its usability (System Usability Scale score of 67.7) and acceptability (4.4 out of 5 in the 'Willingness to Use' category in the relative Technology Acceptance Model). The results from the tests on healthy participants and the feedback from clinicians encourage further exploration of our framework, to verify its potential in supporting arm rehabilitation for individuals with neurological disorders.}




\maketitle

\section{Introduction}\label{sec1}

% INIZIARE SPIEGANDO QUANTO E PERCHE' SIA IMPORTANTE ESEGUIRE ESERCIZI DI RIABILITAZIONE CON PRECISIONE
Executing effective rehabilitation exercises is essential for people suffering from the motor consequences of neurological disorders to regain their lost abilities \citep{Takeuchi2013RehabilitationPlasticity}. The physiological execution of movements reduces their metabolic cost and increases their efficiency. Moreover, performing motor tasks accurately, according to the therapist's specific plan to enhance the activation of the correct muscles and synergies, is fundamental to obtaining positive rehabilitation outcomes and avoiding the occurrence of maladaptive plasticity phenomena \citep{Takeuchi2013RehabilitationPlasticity}.  
During standard rehabilitation, therapists assist, guide and support their patients in executing such tasks, ensuring that they are performed correctly. Augmented feedback on patient performance, which includes input from the therapist, is crucial. It indeed supplements the lack of intrinsic feedback that people often experience following a traumatic event, such as a stroke \citep{Thikey2012AugmentedTrial}, and ensures that they receive real-time corrections for the accurate execution of movements.

When coming to technologically-aided rehabilitation, visual augmented feedback is the most common type of concurrent feedback for upper-limb rehabilitation exercises and has proven to be effective in enhancing patients' kinematic performance \citep{Sigrist2013AugmentedReview}. It includes providing visual indications about the task to be executed or the trajectory to be followed through the use of displays, headsets or projectors \citep{Sigrist2013AugmentedReview}. In such a way, it encourages a precise kinematic execution of the task. At the same time, visual indication of the patient's kinematic performance can help the therapist supervising the therapy to understand the condition of their patients. 

Augmented Reality-based rehabilitation systems can take this a step further. They have demonstrated their potential in offering visual feedback and supporting patients during arm training \citep{Phan2022EffectivenessMeta-Analysis}. Augmented Reality (AR) integrates virtual objects with the real environment surrounding the user, guaranteeing a heightened feeling of embodiment if compared to similar Virtual Reality (VR) applications \citep{Genay2021}. 
The introduction of AR applications into rehabilitation sessions can enable patients to engage in exercises with the guidance of virtual objects. Such objects furnish reliable feedback on how to execute the tasks \citep{Sigrist2013AugmentedReview, Phan2022EffectivenessMeta-Analysis}. AR feedback can be conveyed through various systems, such as displays or wearable headsets. It has been seen that elderly people appreciate head-mounted displays more than screens as they feel more involved in the exercise \citep{Jordan2011AugmentedStroke}. 
% Increasing evidence suggests that the use of augmented visual feedback could be a useful approach to stroke rehabilitation. In current clinical practice, visual feedback of movement performance is often limited to the use of mirrors or video. However, neither approach is optimal since cognitive and self-image issues can distract or distress patients and their movement can be obscured by clothing or limited viewpoints. Three-dimensional motion capture has the potential to provide accurate kinematic data required for objective assessment and feedback in the clinical environment. 
% Feedback on performance plays a central role in skill acquisition. After a stroke, intrinsic feedback mechanisms are often impaired so extrinsic (or augmented) feedback is of great importance for motor relearning. --> https://www.ncbi.nlm.nih.gov/pmc/articles/PMC3541174/

% LACK OF SYSTEMS FOR 3D FEEDBACK ABOUT THE TRAJECTORIES TO BE EXECUTES

% PURPOSE
This is why, in this paper, we present an AR rehabilitation system designed to provide 3D guidance to the users on the trajectories to be executed during training sessions and to evaluate, in real-time, their kinematic performances. Employing the AR headset HoloLens 2 by Microsoft, our system projects holograms of prerecorded trajectories and tracks users' hand movements. These holograms are superimposed onto the real clinical or home environment, providing precise indications for the patient to follow and still allowing for real-object interaction. Our system’s real-time hand-tracking feature includes a colour-coded indication of the user’s precision in following the trajectory, with three different dimensions representing distinct confidence intervals.
We tested our system with 15 healthy individuals to demonstrate that, with the AR system, they can execute the exercises with higher accuracy. Participants performed tests with 4 different exercises with specific trajectories, always wearing the upper-limb exoskeleton AGREE to be used as an instrument to validate the hand-tracking system, recording the real-time positions of the arm of the user.
Additionally, 12 clinicians from a neurorehabilitation centre in Milan evaluated our system in terms of usability and acceptability according to its capacity to fulfil their need for improved kinematic feedback about the patient during rehabilitation sessions, especially when a robotic device is involved.
\section{Related works: Augmented Reality feedback for upper-limb rehabilitation}\label{sec2}

\subsection{The importance of visual feedback}
When dealing with motor learning, extrinsic feedback plays a fundamental role in improving performances \citep{Thikey2012AugmentedTrial}. 
Providing feedback is crucial to ensure the effectiveness of arm rehabilitation tasks, especially given the particular complexity of physiological upper-limb movements \citep{Sigrist2013AugmentedReview}. 
There are different ways to give feedback about the patient's performance during a rehabilitation exercise, including visual, auditory, haptic or multimodal, which is a combination of the different methods. Visual is the most investigated modality for feedback and its efficacy has been demonstrated, especially for complex tasks \citep{Sigrist2013AugmentedReview}.\\
% EXAMPLES OF VISUAL FEEDBACK SYSTEMS
% just 2d
Mirror visual feedback, showing movements of the contralateral arm, can help improve the control abilities and coordination in post-stroke patients \citep{Urra2015TheUpper-limb, Kim2023UtilizationMeta-Analysis}.
Visual representations of abstract variables representing the patients' performances can help them adjust and improve their motor behaviour. For instance, bar plots indicating inter-limb displacement can help in coordination tasks \citep{Maslovat2009FeedbackHypothesis}, force plots can indicate how close the subject is to the target in manual exercises \citep{Snodgrass2010Real-timeMobilisation} and tables with embedded LEDs or monitors can guide to reach target positions during rehabilitation tasks \citep{DallaGasperina2023AGREE:Patients} or inform about how precisely the patient is executing the movement \citep{Simonsen2017DesignMovement}.
All the aforementioned options have many advantages but lack tridimensionality, which is fundamental to represent the complexity of multidimensional arm movements \citep{Sigrist2013AugmentedReview}.

In this sense, a virtual-reality-based system can offer an effective solution. Making the patient observe a screen with a concurrent virtual reproduction of the task they are executing, such as ball-hitting with a racket \citep{Todorov1997AugmentedTask}, is an efficient option to train specific motor gestures. Headset applications can offer a more immersive 3D experience. For example, the system described by Semblantes \textit{et al.} \citep{Semblantes2018VisualPatients} tracks and reproduces in a virtual environment the gestures of the patient, wearing an Oculus. It then delivers terminal feedback by comparing the patient's exercises with gestures executed by therapists.
Visual-feedback frameworks can be helpful not just for the patients, but also for the clinicians that supervise them, especially in the context of robotic rehabilitation. In fact, therapists and doctors lose perception of the condition of the treated limb when patients wear robotic devices, which strongly reduces the acceptability level towards rehabilitation technologies \citep{Laparidou2021PatientMeta-synthesis, Luciani2023TechnologyPerspectives}. 
In this sense, the system proposed by Quintana \textit{et al.} \citep{Quintana2022AuthoringExercises} is an example that merges virtual reality games for rehabilitation with an interface for real-time supervision of the patient. This interface is designed to help the therapist better gain insight into the rehabilitation process.
Despite all the advantages represented by the application of VR systems for rehabilitation, virtual reality carries some disadvantages. Immersive VR systems, such as headsets, may cause feelings of dizziness and nausea in some people who wear them \citep{Saredakis2020FactorsMeta-Analysis}. This is especially due to the total isolation from the external world, which may cause discrepancies between what the users see and what they perceive with the other senses. Moreover, VR systems do not provide chances for interaction with the real environment surrounding the patient or object manipulation \citep{Alamri2010AR-REHAB:Rehabilitation}.
AR systems can provide an effective answer to these limitations.

\subsection{AR for upper-limb rehabilitation}
AR technologies can be of great support in the rehabilitation field, providing interactive solutions that enhance users' active participation in the therapy \citep{Phan2022EffectivenessMeta-Analysis}. Game-based applications have been developed using AR to let patients exercise their arms, following indications or playing with virtual elements while keeping a feeling of embodiment in the real world and interacting with objects. 
De Crignis \textit{et al.} \citep{deCrignis2023RoboticStudy} demonstrated the feasibility of the application of serious AR games in clinical environments. They tested a 3D puzzle that consisted of constructing a house with some bricks collecting positive feedback from both therapists and patients.
The work presented by Bur \textit{et al.} \citep{Bur2010AugmentedRehabilitation}\citep{Bur2010AugmentedRehabilitation} implemented different AR applications that exploit some markers to transfer a real-life cube into the virtual environment. Such applications included a brick-breaker game and a task-oriented game where the user has to move the cube in different required locations. The AR system evaluates the marker position and indicates if the exercises are performed in the right way.
De Assis \textit{et al.} \citep{Assis2014AnStudy} proposed an innovative application that uses a virtual arm on the screen to replace an impaired limb for people undergoing post-stroke rehabilitation therapy, to help them gain shoulder range of motion/movement. They demonstrated that the integration of their game into post-stroke rehabilitation therapy improved the outcomes of the treatment.
\\Virtual objects can provide accurate feedback to the patients, enhancing their participation in the rehabilitation sessions \citep{Sigrist2013AugmentedReview, Phan2022EffectivenessMeta-Analysis}. AR can be used to measure the speed and direction of upper-limb movements and to provide feedback about the patient's capability to adapt and avoid obstacles \citep{Bank2018Patient-TailoredStroke}. These indications are of key importance also for the therapist supervising the training sessions.\\
From all these examples we can see how AR has the potential to support and inform rehabilitation sessions in an effective and engaging way. Despite the presence of different AR applications for rehabilitation, to the best of our knowledge, no previous work implemented an AR-based feedback system for 3D tracking of the hand of the user during custom rehabilitation exercises, with real-time performance indication.
% \subsection{AR-based visual feedback systems}
% XXX FUSING NECESSITY OF FEEDBACK WITH EFFICACY OF AR: FB systems
% VISUAL FEEDBACK TO GUIDE THE USER OR PROVIDE TIMING
% TARGETS -- AGREE currently limited
% INDICATION OF A PERFORMANCE VARIABLE

% LIMITATION OF THE 2D OPTIONS
% WHY AR HEADSET AND NOT SCREEN

\section{Materials}\label{sec3}
\subsection{HoloLens 2 and correlated tools}
\begin{figure}[h!]
    \centering
    \includegraphics[width=0.4\linewidth]{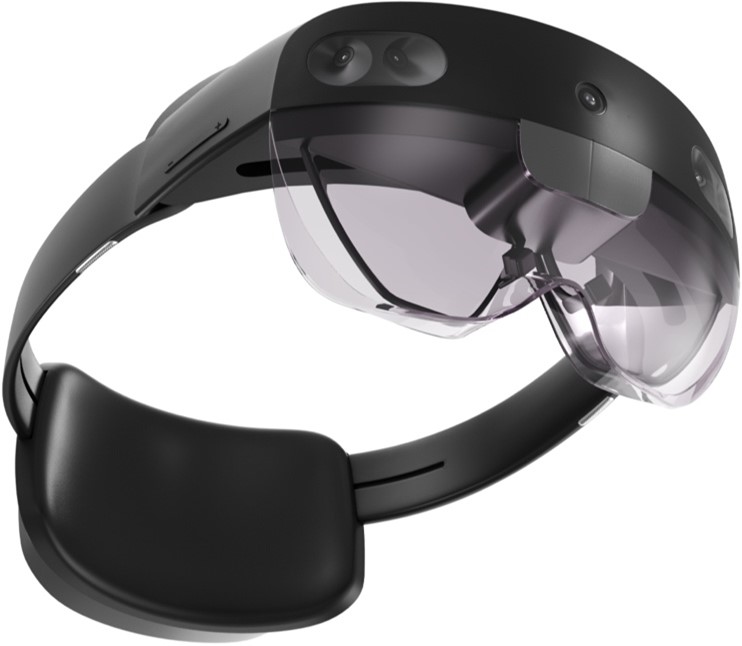}
    \caption{HoloLens 2 Headset \citep{MicrosoftHoloLens2024}}
    \label{fig:holo}
\end{figure}

The HoloLens 2 headset (Microsoft, Washington, USA), depicted in Fig.\ref{fig:holo}, projects virtual images onto a screen through prisms and optical systems, precisely directing the images over an appropriately expanded area \citep{MicrosoftHoloLens2024}. 
In such a way, the wearer of the headset sees the virtual objects superimposed on the real environment. The interaction between the user and the virtual space is made possible through the hand-tracking algorithm provided by the headset itself, which recognizes the location of the hand and its segments in a referenced space. The hand-tracking enables the execution of various actions such as pinching or clicking. 
HoloLens 2 is equipped with numerous sensors, such as Inertial Measurement Units (IMUs), infrared and RGB cameras and depth-sensing cameras. It is completely wireless and weighs only 556g. The operating system is a trimmed-down and adapted version of Windows 10 for virtual inputs and includes a comprehensive and open-source development environment \citep{MicrosoftHoloLens2024}.

To realise custom holograms for HoloLens and program our virtual application, we employed the cross-platform game engine Unity, by Unity Technologies \citep{Unity2024}. We integrated several external libraries to enhance the Unity environment with specific functionalities, such as:
\begin{itemize}
    \item \textit{Windows Mixed Reality} (WMR), to make the Unity development environment compatible with the HoloLens and handle interactions between hands and three-dimensional objects within the scene.
    \item \textit{Tuberenderer} by "Sixth sensor" \citep{TubeRenderer2024}, to enable the creation of tubes around complex and nonlinear objects and create our custom trajectories.
\end{itemize} Unity programming language is $C\#$, and the developer platform we used is Visual Studio. Also, Shapr3D, a 3D CAD software, was employed to create complex 3D models and then import them into the Unity environment \citep{Shapr3D2024}.

\subsection{AGREE}
For this work, we needed a reliable arm-tracking system, to evaluate the kinematics of the different joints and the hand of the users while executing rehabilitation exercises, both with and without the use of the headset. The AGREE exoskeleton \citep{DallaGasperina2023AGREE:Patients}, with its relative encoders, provides real-time indication of the angular position of its joints (running with a 5 kHz sampling frequency). Thanks to the robot's kinematic model, the end-effector's position is then easily calculated.
AGREE is a four-degrees-of-freedom upper-limb exoskeleton by Politecnico di Milano, designed to assist subjects suffering from motor deficits caused by neurological disorders in their rehabilitation process.
Using AGREE as an evaluation tool for kinematic performance, while using our application, also offers the chance to test the AR framework in a realistic condition of use within a robotic rehabilitation session.

\section{Methods: implementation of the application and testing}\label{sec4}
\subsection{Application structure}

The application we developed for HoloLens 2 has been thought to include three main phases:
\subsubsection*{(i) Calibration}
\begin{figure}[h!]
    \centering
    \includegraphics[width=1\linewidth]{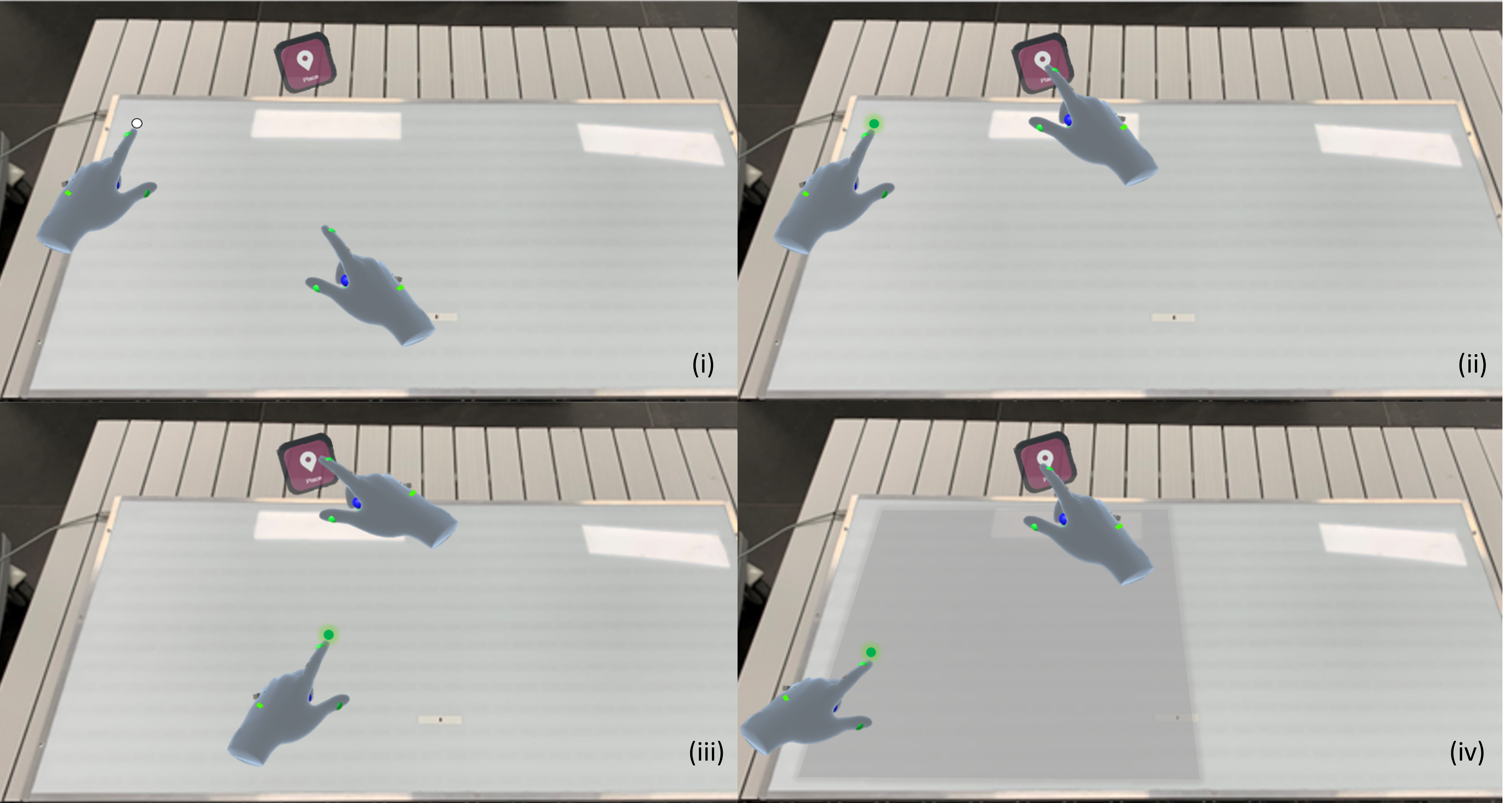}
    \caption{Representation of the calibration procedure: (i) A white sphere appears and the user can locate it where they prefer, (ii) Pressing the \textit{Place} button, the sphere turns green and keeps its position, indicating the first reference point, (iii) A second reference point is located in the same way, (iv) After the third point is located, the relative plane appears in the space.}
    \label{fig:calib}
\end{figure}

This phase is needed to establish a correspondence between the virtual local reference system and the real workspace. It is crucial to ensure that the user can execute the tasks with the correct spatial feedback. It is thought to be executed by a technician or therapist, to avoid any fatiguing gesture for the patient. The implemented method employs the hand-tracking algorithm of HoloLens to locate three reference points, defining a plane, in space. When the calibration begins: 
\begin{itemize}
    \item A white sphere appears on the tip of the left-hand index finger of the user to indicate the virtual marker that needs to be fixed;
    \item The user can move the left hand and point the index finger with the sphere in the location they prefer;
    \item Clicking the "place" button that is present on the scene with the right hand, the user can fix the marker's position and the white sphere turns green.
\end{itemize} 
These three steps need to be repeated three times to build a Cartesian reference and the working plane. We always used a table to help us in the definition of the reference plane.
The whole procedure is represented in Fig.\ref{fig:calib}.

At the end of the Calibration phase, the therapist is offered the possibility of recording a new custom trajectory. Employing the cameras of the headset, our system can track and record the movements of the hand of the clinicians and create trajectories to be then proposed to the patients.

\subsubsection*{(ii) Trajectory selection and manipulation}
\begin{figure}[h!]
    \centering
    \includegraphics[width=1\linewidth]{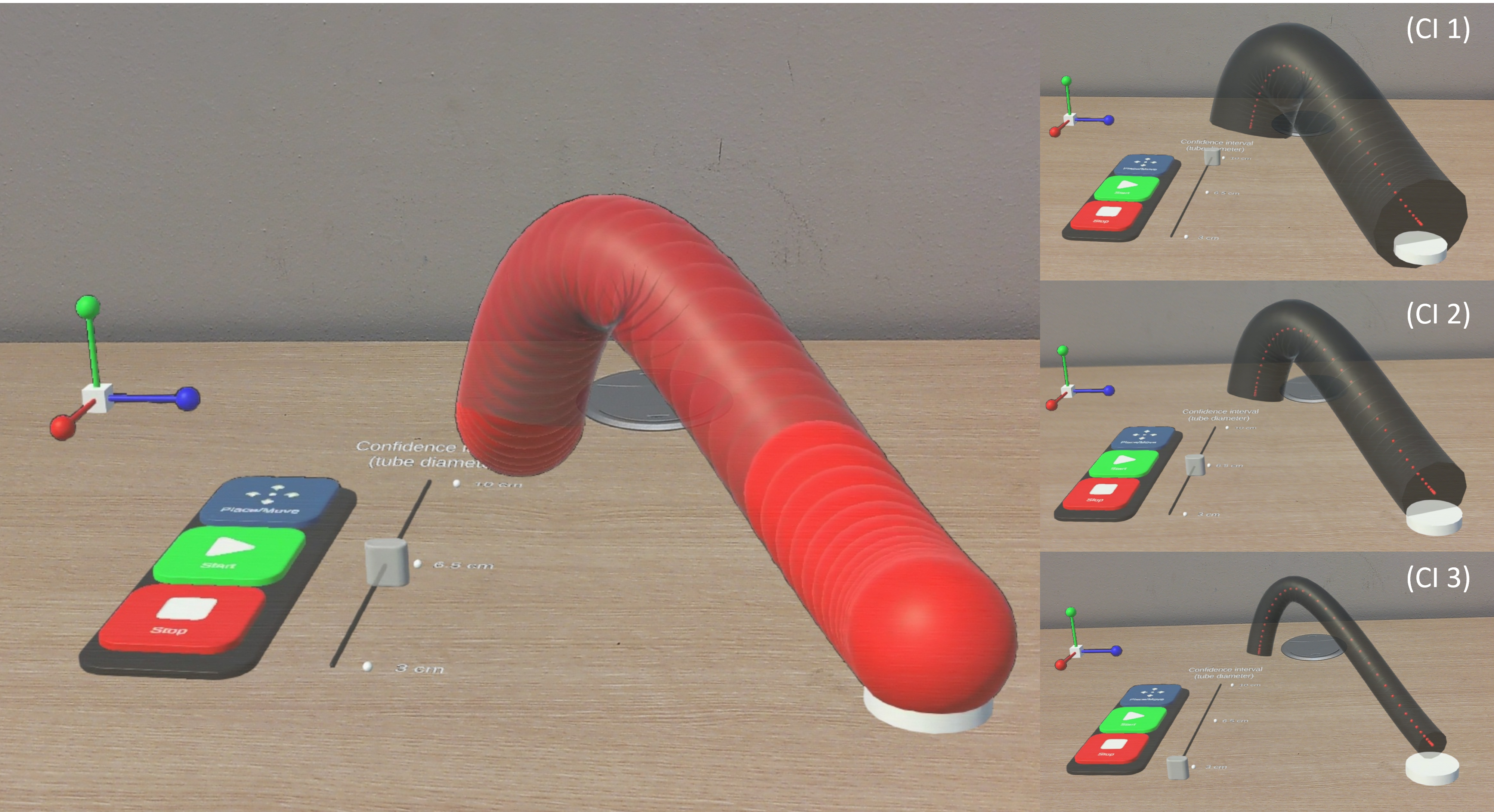}
    \caption{Trajectories superimposed to the real environment. (LEFT) The appearance of a trajectory tunnel before the beginning of the execution, with the functioning buttons and the slider. The tunnel appears completely red before the user starts moving. (RIGHT) The pictures show the appearance of the trajectory with the three confidence intervals (CIs).}
    \label{fig:trajectories_app}
\end{figure}

This phase allows the visualisation of the trajectories and their manipulation in the space. Once the calibration is completed, the patient can wear the headset and sit in front of the table where the virtual workspace has been located. The trajectories to be used for the rehabilitation session are displayed in front of the user, on the virtual plane. As seen in Fig.\ref{fig:trajectories_app}, each trajectory has a white base, that can be clicked to select the trajectory, and a series of three associated buttons and a slider. 

The blue \textit{Place/Move} button can be selected to move the trajectory. By clicking it with the left hand, the user can then use the right hand to move the base of the trajectory and locate it in correspondence to the desired starting point. A second click is needed to fix such a position and disable the motion. Since the horizontal plane has been defined during the calibration, the movement of the trajectory is allowed only along the coordinates of such plane but will keep the vertical position fixed at the level of the plane.
The slider, positioned to the right of the buttons, is needed to select the desired confidence interval (CI) for the trajectory. We have three possible trajectory diameters ($10cm$ - $6.5cm$ - $3cm$), representing how precise we want the user to be while executing the task. A CI of $10cm$ means that we allow the patient to deviate from the desired trajectory for a maximum of $5cm$ without giving them any negative feedback. With a CI of $3cm$, on the other side, we require them to be as precise as possible and will identify as an error any deviation that is bigger than $1.5cm$ from the centre of the tunnel. The appearance of a trajectory with different CIs can be seen in Fig.\ref{fig:trajectories_app}.
The green \textit{Start} button has to be clicked at the beginning of the execution of the task. The trajectory changes its appearance and results as composed of red spheres with a diameter that coincides with the selected CI.
The red \textit{Stop} button, finally, needs to be clicked at the end of the exercise.
For patients unable to control the contralateral arm, the buttons can be selected by the therapists via a computer that communicates to the HoloLens through its Holographic Remoting application.

\subsubsection*{(iii) Execution and Scoring}
\begin{figure}[h!]
    \centering
    \includegraphics[width=1.0\linewidth]{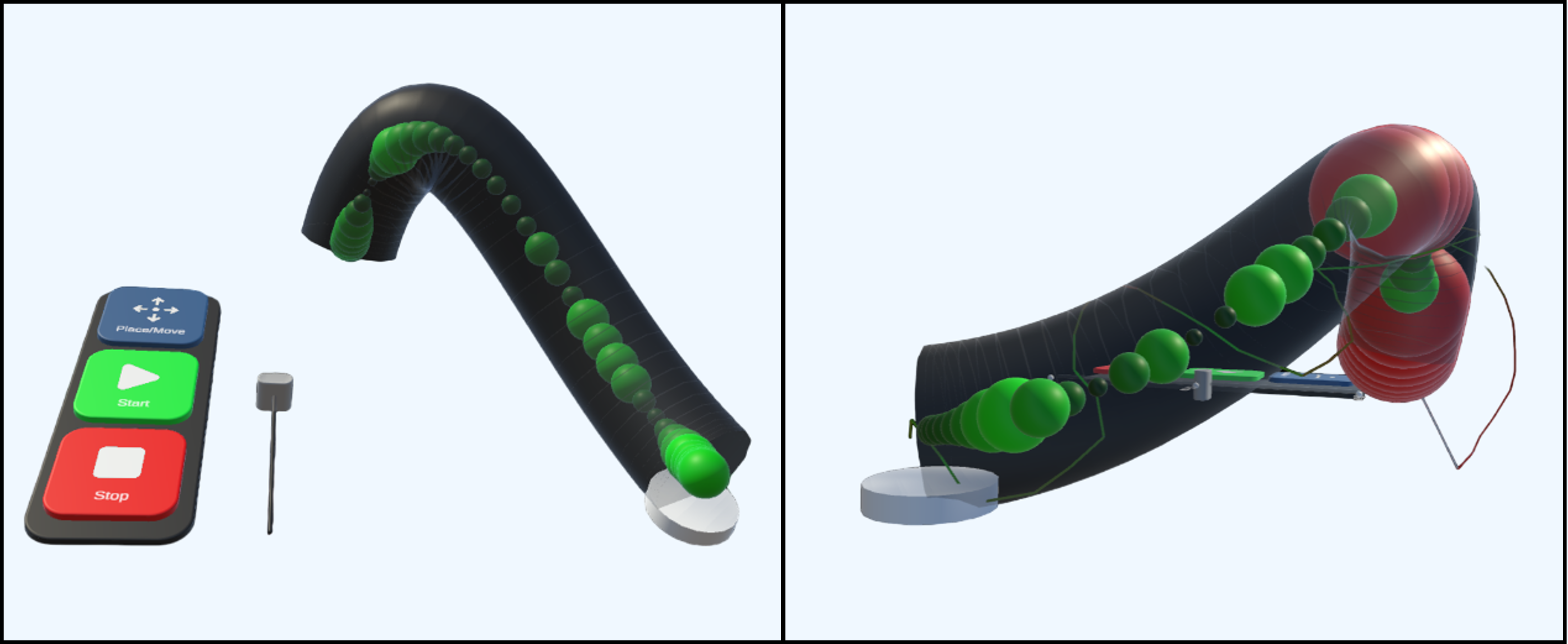}
    \caption{Feedback system showing the error-based feedback system. (LEFT) The spheres have different sizes and colours according to the error in their specific position along the trajectory. (RIGHT) At the end of the task, the system provides a snapshot of the movements executed, represented by a line that tracks the whole path of the hand.}
    \label{fig:error_rep}
\end{figure}

Once the \textit{Start} button has been clicked, the user can execute the task trying to follow the visual feedback represented by the trajectory. While they move their hand, the HoloLens tracking system instantaneously computes the position of the centroid of the hand and compares it to the position of the closest via-point of the trajectory. The distance is defined as a simple Euclidean difference.
The spheres composing the trajectory change their appearance in real time, modifying their size according to the computed error. They "deflate" and change colour proportionally to the proximity between the hand and the trajectory's centre: the smaller the error, the smaller and the greener the sphere. 
If the error is bigger than the CI, the spheres remain red. If the centroid crosses exactly the desired path, the spheres are the smallest possible and coloured in dark green. We developed two options for the real-time error feedback system. With the first one, the feedback can be "overwritten" during every repetition of the movement, so that the user can notice the improvements by seeing the spheres scaling down. With the second option, the trajectory returns completely red after every repetition.
In Fig.\ref{fig:error_rep} we can appreciate what happens during the execution of the task, thanks to the real-time error feedback system. In the left picture, we can appreciate the different sizes and colours of the spheres. The right picture represents what is visible after the end of the task execution. The thin coloured line represents the actual path followed by the user.

\begin{figure}[h!]
    \centering
    \includegraphics[width=1\linewidth]{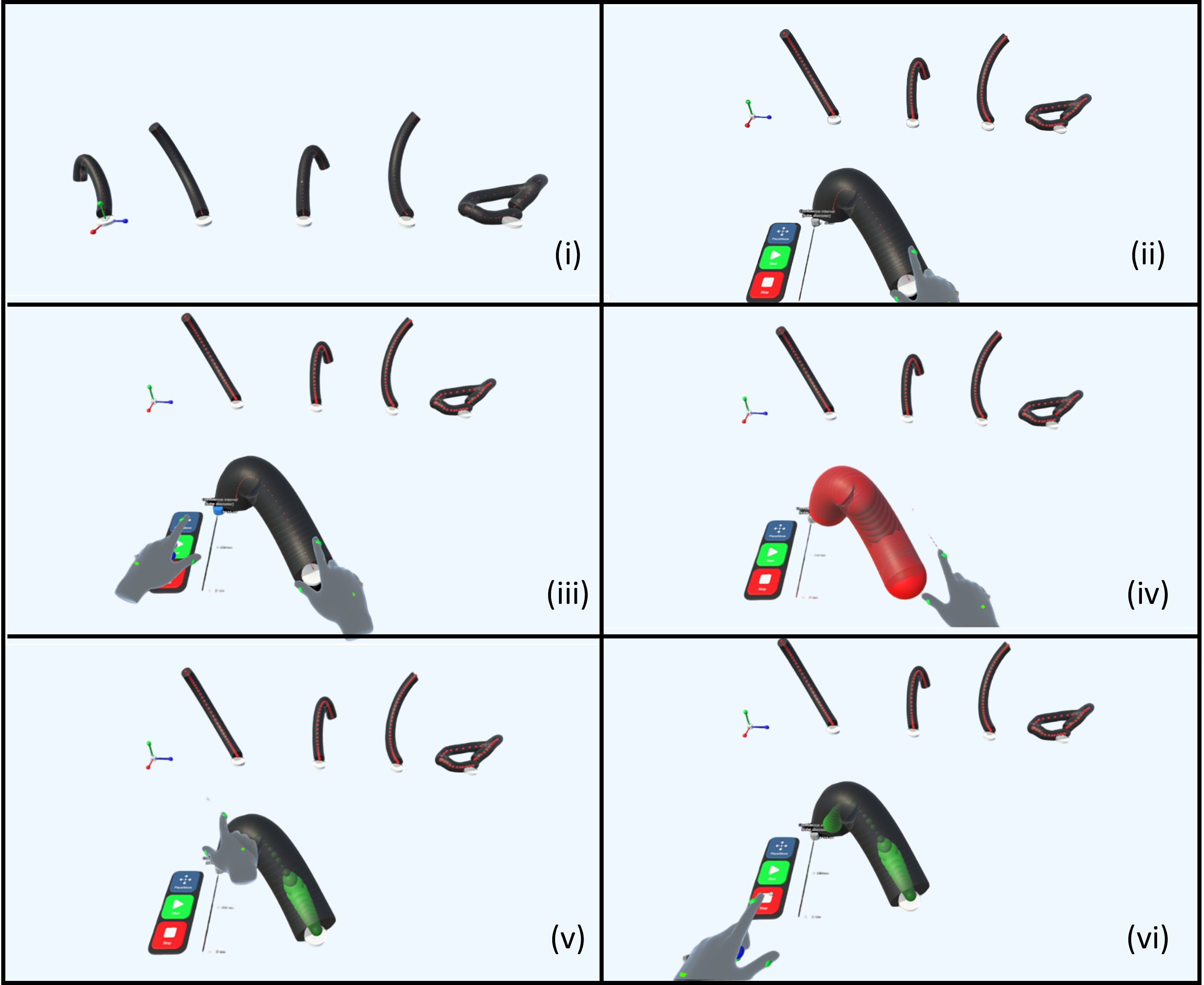}
    \caption{Steps to follow to execute a task with the HoloLens feedback: (i) Select the trajectory from the options; (ii) Position the trajectory on the Start Point using the blue button; (iii) Start the exercise with the Green button; (iv) Start moving trying to keep the hand in the red tunnel; (v) Execute the trajectory: the tunnel turns green if the execution is correct; (vi) Press the red button to stop.}
    \label{fig:steps}
\end{figure}

%\subsubsection*{(iv) Scoring}
%From the beginning of the exercise, along with all the possible repetitions of the task, the program calculates the distance between all the points of the trajectory and the centroid of the right hand, saving them within an array. As the hand moves, the distances change, and the array is updated by replacing the values of the indexes only if the distance calculated in the current iteration is lower than the one contained in the array. Consequently, at the end of the execution, the resulting error array will consist of the absolute minimum distances. Immediately after pressing the 'stop' button, a panel is displayed in front of the subject. Such a panel reports a histogram showing the error array. Using the hand as a pointer and hovering over individual bars, the numerical value represented by each bar is displayed (see Fig.\ref{fig:error_panel}).
%\begin{figure}[h!]
%    \centering
%    \includegraphics[width=0.75\linewidth]{figures/error_panel.png}
%    \caption{Error panel summarizing the execution error along the repetitions of the exercise.}
%    \label{fig:error_panel}
%\end{figure}
Fig. \ref{fig:steps} summarizes the steps to be followed by the user for a single trajectory and one CI.

\subsection{Testing protocol: efficacy with healthy subjects}
\begin{figure}[h!]
    \centering
\includegraphics[width=0.5\linewidth]{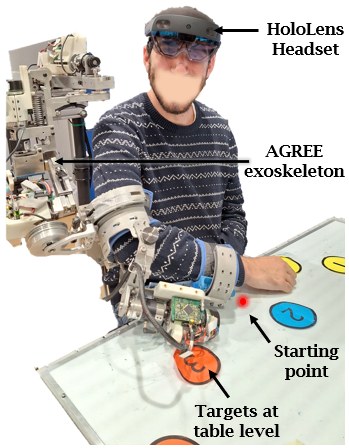}
    \caption{Experimental setup with a subject wearing both the exoskeleton AGREE and the headset. In front of the subject, we have a table with targets to be used as reference points for the tasks.}
    \label{fig:setup}
\end{figure}

We recruited 15 healthy subjects with an age between 24 and 32. We asked them to sit wearing AGREE and the HoloLens 2 on their head. In front of the subject, we positioned a table with some targets on it, at different height levels. Figure \ref{fig:setup} shows the experimental setup.
At the beginning of each subject's session, they were instructed about the use of the two systems and they had a moment to familiarize themselves with the use of AR.
Before asking for their active participation, we executed the calibration phase and defined the starting point for the first trajectory. 
The exercises proposed to the users, developed to be proposed during rehabilitation sessions, were four: (T1) Reaching a point at the table level towards the left of the subject; (T2) Reaching a point at the shoulder level towards the left of the subject; (T3) Reaching a point at the shoulder level towards the front of the subject; (T4) Draw a clockwise circle at table level, starting from the starting point.
The participants to the experiments signed an informed declaration of consent and the tests were approved by the Ethical Committee of Politecnico di Milano (approval n. 8/2022 - 16/02/2022).

\subsubsection*{Exercises execution}

For each exercise, participants were initially instructed about the movement to be executed.
They were then asked to follow some steps and repeat the exercise in different conditions.\\
Without wearing the headset:
\begin{itemize}
    \item Perform the movement back and forth, having as references just the start and end points of the trajectory;
    \item Repeat the movement five times.
\end{itemize}
Wearing the HoloLens 2:
\begin{itemize}
    \item Select the correct hologram for the trajectory to be executed from the virtual environment;
    \item Perform the corresponding movement back and forth, selecting the bigger confidence interval (C1);
    \item Perform the corresponding movement back and forth, selecting the medium confidence interval (C2);
    \item Perform the corresponding movement back and forth, selecting the smaller confidence interval (C3).
\end{itemize}
The order of application of the different exercise conditions was randomised to avoid the influence of any learning effect on the global results.

\subsubsection*{Error evaluation}
To guarantee the repeatability of the results, comparing the kinematic error both with and without the use of HoloLens, our participants always wore AGREE. AGREE records the real-time joint positions of the four joints and, from them, the end-effector position is computed. At the end of each exercise, the five repetitions are isolated and normalised in time to be compared with the desired trajectory, which is stored in the trajectory generation system of AGREE \citep{Gasperina2022}. 
For every time sample $i_th$ of each repetition, the formula used to evaluate the kinematic error in the end-effector space uses the root mean squared error (RMSE):
\begin{equation}
    RMSE^{i} = \sqrt{(x^{i}-x_d^i)^2+(y^{i}-y_d^i)^2 + (z^{i}-z_d^i)^2}
\end{equation}
where $x^{i}, y^{i}, z^{i}$ are the actual positions along the analysed repetition, while  $x_d^{i}, y_d^{i}, z_d^{i}$ are the desired ones.
The error is then mediated over the single repetition, and over the five repetitions, giving the final result:
\begin{equation}
    ERR_s^{T_j} = \frac{1}{N}\sum_{n=1}^N\left(\frac{1}{I}\sum_{i=1}^IRMSE^{i,n}\right)
\end{equation}
where $s$ is the participant analysed, $T_j$ is the exercise, $I$ is the total number of time samples for the normalised repetitions and $N$ is the number of repetitions (5). In joint space, the error is first evaluated joint by joint and then averaged.
This is repeated for every subject and exercise. 

The errors have been gathered per exercise and then, globally, per confidence interval. We statistically analysed them to compare the results obtained with HoloLens, distinguishing the three CIs, with those obtained without the visual feedback guaranteed by our AR system.
First, we checked the non-parametricity of our data distributions using the Kolmogorov-Smirnov test.
We then created a generalized linear mixed (GLM) model with two factors - Exercise and Condition - with four levels each (T1, T2, T3 and T4 for Exercise; without HoloLens, C1, C2 and C3 for Condition) to highlight the effects of these factors on the error. The model considers Exercise and Conditions as fixed effects and includes participants as a random effect.
After identifying significant effects, we used the Wilcoxon signed-rank test to compare the errors obtained in the various conditions. All statistical analyses were done using MATLAB 2022b.

Due to the intrinsic kinematic redundancy of the human arm, it is possible to move the hand along the same trajectory while positioning the elbow and shoulder joints in different configurations. To confirm that our end-effector-based feedback system does not lead to joint misalignments or introduce increased errors in joint space, the analyses and comparisons performed for the error in the end-effector space, were then repeated for all the joints.

\subsubsection*{Questionnaire}
At the end of the protocol, the participants were asked to complete a questionnaire (see Appendix \ref{AppA}). The questionnaire reported 22 sentences, regarding the intuitiveness and usability of our application. 
The sentences were classified into five categories: (i) Clarity of the calibration; (ii) Usability of the trajectory manipulation mechanism; (iii) Effectiveness of the guiding interface; (iv) Reliability and stability of the hand-tracking system (v) Comfort of the device.
The volunteers had to indicate their degree of agreement with a score on a Likert scale from 1 (strongly disagree) to 5 (strongly agree).

\subsection{Therapists' acceptability and usability evaluation}
We presented our system to 12 professionals working in the neurorehabilitation clinic \textit{Casa di Cura Igea S.p.A.}, in Milan. After a 30-minute session of training and testing of the device, we asked them to complete two questionnaires (see Appendix \ref{secA2}) and evaluate the system in terms of usability and acceptability. During the tests, they experimented with different trajectories and confidence intervals, replicating the experimental iterations proposed to the healthy subjects. Additionally, they tested the various button functionalities and tried out the additional option that we developed for the therapist's side, which allowed them to record trajectories in real-time from demonstrations using their hands. Table  \ref{tab:ther} reports some personal information about the professionals involved (such as age, gender, profession, years of experience working in the rehabilitation field and attitude towards technology on a scale from 1 to 5).
\begin{table}[h!]
    \centering
     \caption{Registry information about the participants to the evaluation. OT: occupational therapist, PH: physiotherapist, BIO: clinical bioengineer. The Field Attitude Towards Technology Reports the Mean Value of the Scores (On a Scale 1–5) They Attributed to 4 Sentences Indicating Their Level of Interest and Confidence When Dealing With New Technologies and With Technology in Their Everyday Life.}
        \label{tab:ther}
   % \resizebox{\columnwidth}{!}{
%\renewcommand{\arraystretch}{1.3}
    \begin{tabular}{cccccc}
    \toprule
    %\rowcolor{blueP!40}
  {\textbf{}} & \multicolumn{1}{c}{\textbf{Age}} & \multicolumn{1}{c}{\textbf{Gender}} & \multicolumn{1}{c}{\textbf{Profession}} & 
  \multicolumn{1}{c}{\textbf{Years of Work}} &
  \multicolumn{1}{c}{\textbf{\begin{tabular}[c]{@{}c@{}}Attitude Towards\\ Technology ( /5)\end{tabular}}} \\ \midrule
\textbf{S.1}    & 32                               & M                                   & PH                                    &4   & 5                                                                                                        \\
\textbf{S.2}    & 26                               & F                                   & OT                              &2         & 3.25                                                                                                     \\
\textbf{S.3}    & 34                               & F                                   & PH                            &10           & 4.5                                                                                                      \\
\textbf{S.4}    & 24                               & M                                   & OT                            &4           & 4.5                                                                                                      \\
\textbf{S.5}    & 32                               & F                                   & PH                           &10            & 4.25                                                                                                     \\
\textbf{S.6}    & 37                               & M                                   & OT                          &15             & 3.75                                                                                                     \\
\textbf{S.7}    & 37                               & F                                   & PH                             &14          & 4                                                                                                        \\
\textbf{S.8}    & 37                               & F                                   & OT                            &14           & 3.25                                                                                                     \\
\textbf{S.9}    & 31                               & F                                   & BIO                           &4           & 5                                                                                                        \\
\textbf{S.10}   & 33                               & M                                   & PH                              &9         & 5                                                                                                        \\
\textbf{S.11}   & 31                               & M                                   & PH                            &5           & 3.25                                                                                                     \\
\textbf{S.12}   & 41                               & M                                   & PH                               &16        & 4.5     \\ \bottomrule                                 
    \end{tabular}%}
\end{table}
For the usability part, we presented the 10 classical questions for a System Usability Scale (SUS) evaluation \citep{Bangor2008AnScale}. After the completion of the questionnaire, we calculated the SUS score, for every participant, 
consistently with the literature. A score of 100 represents the best possible usability, while 68 indicates the threshold for an acceptable usability level. To inspect the acceptability of the system, we presented 21 questions, divided into 6 categories, and evaluated the answers to build a Technology Acceptance Model (TAM) \citep{Marangunic2015Technology2013}. We considered the three classic TAM categories (\textit{Willingness To Use} (WTU), \textit{Perceived Usefulness} (PU), \textit{Perceived Ease of Use} (PEOU)) but also included three additional categories specifically referred to our system features (\textit{Importance of Tridimensionality} (TRI), \textit{Importance of therapy independence} (IND), \textit{Importance of Graphical aspects} (GRAPH)). After the evaluation of Cronbach's alpha over the collected answers, two questions were eliminated (one for WTU and the other for GRAPH). The correlation between the different categories' results was examined by calculating the Pearson correlation coefficients and constructing a multiple linear regression model to connect the categories.

\section{Results}\label{sec5}
\subsection{Efficacy with healthy subjects}
\subsubsection*{Error analysis results}

Table \ref{tab:error}, displays the mean error computed for all the exercises without and with the HoloLens, for all the CIs and exercises. It also presents the global results (both mean and standard deviation of the error), obtained by averaging the errors across the four different exercises (i.e., T1, T2, T3, T4). 
The analysis of the generalised linear mixed model indicates that both the type of exercise executed (factor Exercise) and the condition for execution (factor Condition) have a significant effect on the error ($p_{value} < 0.05$). On the other side, the effect of the interaction between the two factors is not significant ($p_{value} > 0.05$). This suggests that the difference in scores between conditions is consistent across different exercises.
Since we are concentrating on the differences in errors when using our application versus not using it, and we have determined that the tasks do not affect these variations, the statistical evaluation using the Wilcoxon signed-rank test indicated in Table \ref{tab:error} focuses on the comparison between the \textit{No Hololens} condition and the other three.
\renewcommand{\arraystretch}{1.5}
\begin{table}[h!]
\caption{Error summary, both in joint and end-effector space, for all the feedback modalities applied for the tests (without HoloLens, with HoloLens in the three CIs). *: 0.05-significance / **: 0.01-significance / ***: 0.001-significance levels in the difference between the kinematic error and the results when no HoloLens was used. }
%\resizebox{\columnwidth}{!}{
\footnotesize
\setlength\tabcolsep{2pt}
%\begin{tabular*}{\textwidth}{@{\extracolsep{\fill}} |l|ccccc|ccccc|}
\begin{tabular}{|l|ccccc|ccccc|}
\hline
\rowcolor{blueP!40}
        & \multicolumn{5}{c|}{End-effector space [cm]}                                             & \multicolumn{5}{c|}{Joint space [deg]}                                                      \\
        \rowcolor{blueP!40}
        & T1   & T2   & T3   & T4   & \begin{tabular}[c]{@{}c@{}}Global\\ result\end{tabular} & T1   & T2    & T3    & T4   & \begin{tabular}[c]{@{}c@{}}Global\\ result\end{tabular} \\ \hline
        \hline
No Holo                                    & 2.46 & 2.43 & 2.24 & 1.97 & \textbf{2.28 $\pm$ 0.45}                                    & 9.38 & 11.34 & 10.30 & 7.32 & \textbf{9.58 $\pm$ 0.86}\\ \hline
\hline
Holo C1                                   & 2.23 & 2.28 & 2.11 &  1.63** &\textbf{2.06 $\pm$ 0.57}***                                    & 8.03** & 10.63* & 9.80  & 5.62** & \textbf{8.52 $\pm$ 1.11}***\\
Holo C2                                 &2.19* & 2.22* & 2.16 &1.66** & \textbf{2.06 $\pm$ 0.59}***                                   & 8.29** & 10.09* & 9.39*  & 6.01**  &\textbf{8.44 $\pm$ 1.30}***  \\
Holo C3                                    & 2.19* & 2.33 & 2.14 & 1.68** & \textbf{2.09 $\pm$ 0.56}***                                    & 8.34* & 10.29** &9.03**  & 5.86** & \textbf{8.38$\pm$ 1.20}***\\
\hline
\end{tabular}%}
\label{tab:error}
\end{table}

\subsubsection*{Questionnaire results}
Table \ref{tab:tab_quest} reports the mean and standard deviations of the scores assigned to each of the categories analysed.
\begin{table}[h!]
 \caption{Questionnaire results: mean and standard deviation of the scores for each category, on a scale from 1 (strongly disagree) to 5 (strongly agree).}
    \label{tab:tab_quest}
    \centering
    \begin{tabular}{cc}
    \toprule
        Category & Score \\
        \midrule
         Clarity of the
calibration & $4.04 \pm 0.62$\\
         Usability of the trajectory manipulation mechanism & $4.13 \pm 0.23$ \\
         Effectiveness
of the guiding interface & $4.57 \pm0.37$\\
         Reliability and stability of the hand-tracking system& $3.57 \pm 0.79$\\
         Comfort of the device& $4.45 \pm 0.44$\\
         \bottomrule
    \end{tabular}
\end{table}

\subsection{Therapists' SUS and TAM results}
\begin{figure}[h!]
    \centering
    \includegraphics[width=0.9\linewidth]{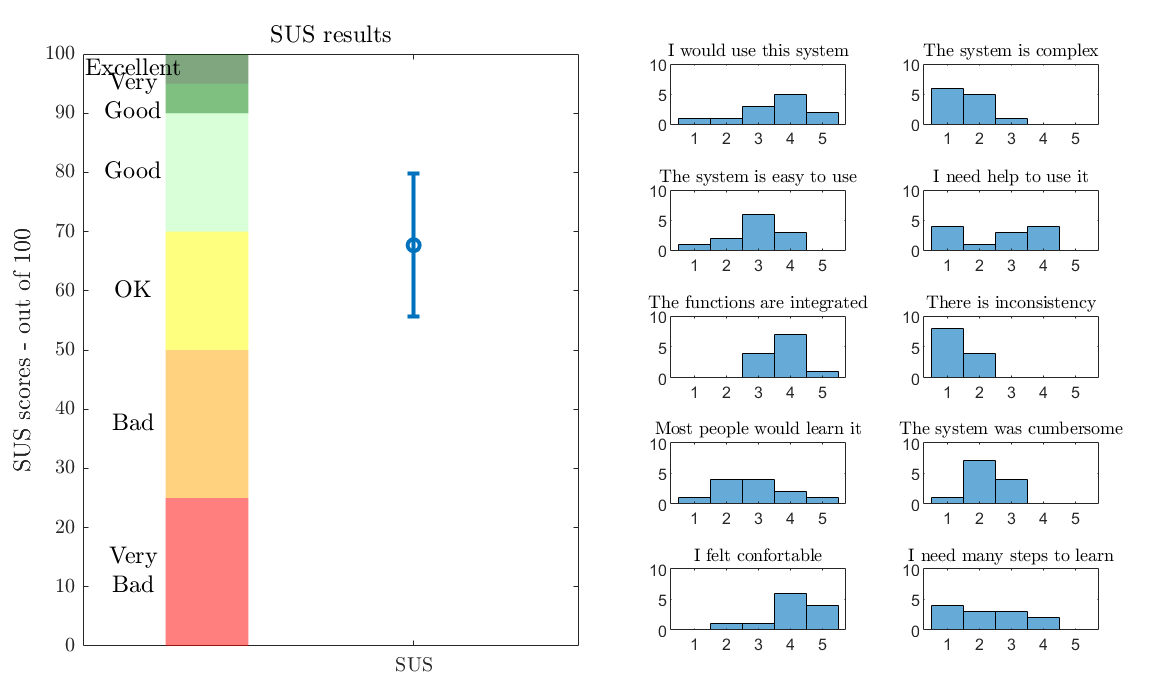}
    \caption{(LEFT) Usability score distribution, with mean and standard deviation, compared to the SUS scale. (RIGHT) Distributions of the scores for the individual SUS questions.}
    \label{fig:sus}
\end{figure}

\begin{figure}[h!]
    \centering
    \includegraphics[width=0.7\linewidth]{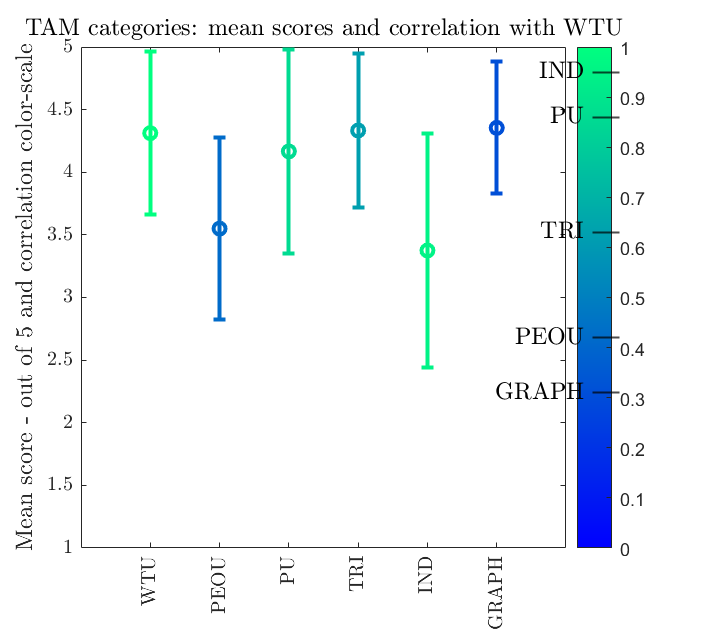}
    \caption{Statistics of the scores (mean and standard deviation) gained by the 6 categories of the TAM model, with a colour-based summary of the correlation between each category and the main WTU category.}
    \label{fig:tam}
\end{figure}

Fig.\ref{fig:sus} represents the usability score distribution in the SUS scale. Our system gained a final SUS evaluation of $67.7 \pm 12.1$ out of $100$. The results of the TAM analysis were mediated after converting the negative items to the proper scale. The mean scores gained by the 6 categories of the TAM model and a colour-based summary of the correlation between each category and the main WTU category are represented in Fig.\ref{fig:tam}. 

Concerning the TAM generated from our questionnaires, we can see how the WTU and PU categories both obtained a mean score of 4.38 out of 5. The Perceived Ease of Use is slightly lower (3.50 out of 5) but has a lower correlation with the Willingness to Use the system (correlation index of 0.42) than the Perceived Usefulness (correlation index of 0.86).  
From the answers we collected, we developed a multiple-regression model to measure the effect of the different variables on the WTU. We considered PU and PEOU to have a direct impact on WTU, while TRI, GRAPH and IND affect the Perceived Usefulness. With a significance level of 0.01, we obtained the following model:
$WTU = 0.73*PU + 0.35*PEOU$, where $PU = 0.59*TRI + 0.50*IND$. $GRAPH$ did not have a significant effect on $PU$.
The $WTU$ scores attributed by the participants and their Attitude Towards Technology are significantly correlated, with a correlation coefficient of $0.63$ ($p_{value} < 0.05$). 

\section{Discussions and Conclusions}\label{sec6}

% MI ASPETTO ANCORA PIU' IMPROVEMENT CON I MALATI SE MIGLIORO CON I SANI, E' POI DA VERIFICARE
 
\begin{figure}[!]
    \centering
    \includegraphics[width=0.4\linewidth]{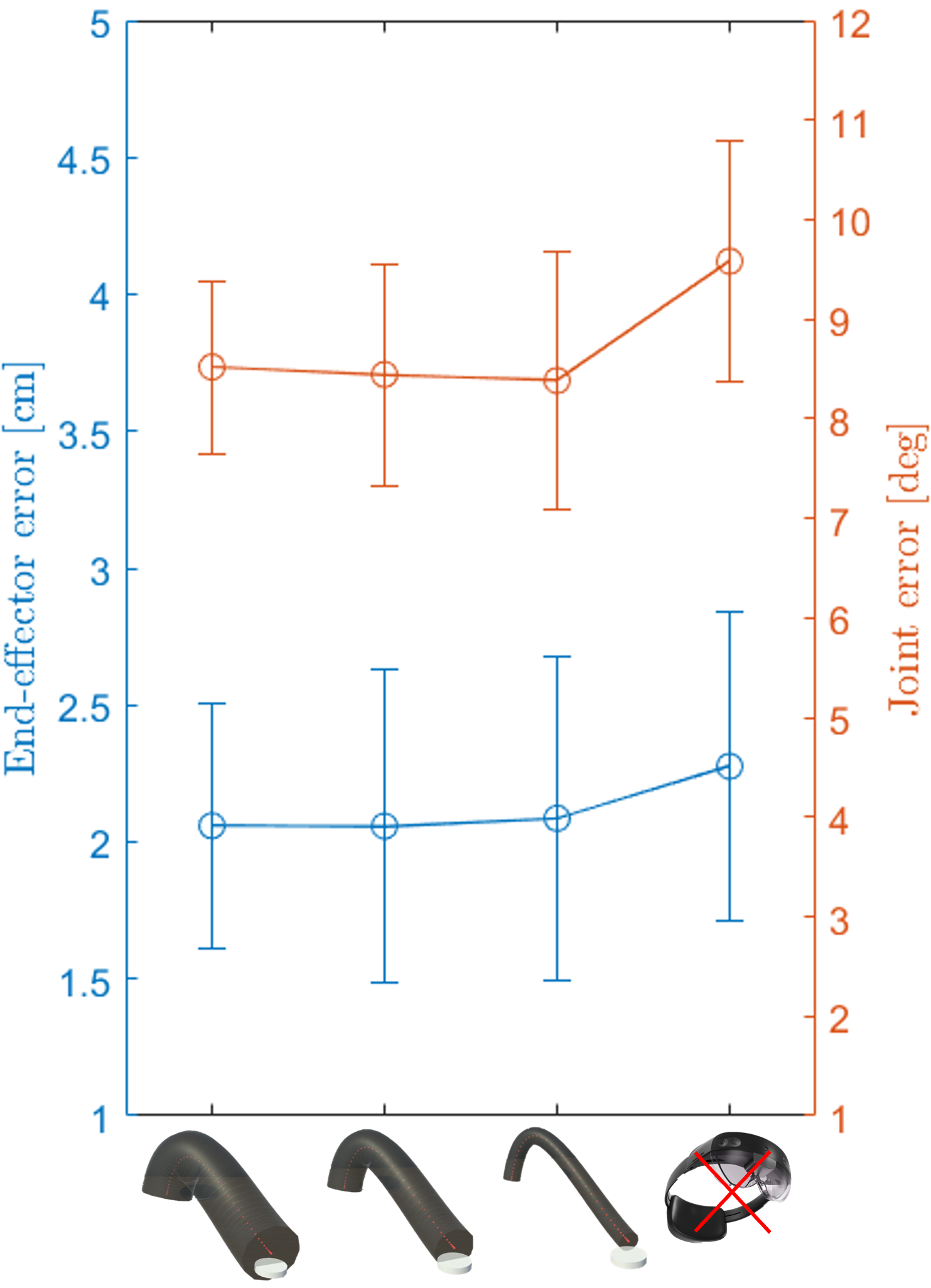}
    \caption{Error statistics in End-effector (left-axis) and Joint (right-axis) space. The results depicted in this summary are evaluated considering all the trajectories and show the median errors collected in C1, C2, C3 and NoHolo modalities (from left to right).}
    \label{fig:stat_result}
\end{figure}

From the results of the GLM model, we observe a statistically significant impact on the error when using our application compared to not using it, and such a difference is not influenced by the the different exercises.
Our feedback system consistently reduces trajectory-execution kinematic errors, as demonstrated in Table \ref{tab:error}, compared to tests conducted without HoloLens. The improvement obtained in the End-effector space is then confirmed in Joint space. The statistical analysis of the results indicates that participants get the greatest improvement when executing T4, thanks to the feedback system. This could be attributed to the fact that the fourth trajectory is more complex and therefore less intuitive to execute. This assumption is generally confirmed if we consider that T2 and T3, which are more linear than the other two trajectories, result in slightly smaller error reductions. 
In Fig.\ref{fig:stat_result} we can appreciate the global trend of the errors along with the different feedback modalities, both in End-effector and Joint space.
If no big difference is observed between the errors obtained with the three CIs in the end-effector space, C3 yields the best results in the joint space. We could observe that, during the execution of the tasks with the feedback system showing the smallest confidence interval, the users moved slowly (time for cycle execution in C3 is $ t_{C3} =6.07\pm0.85 s$ against $ t_{C1} =4.69\pm0.71 s$ and $ t_{C2} =4.99\pm0.77 s$), trying to carefully track the trajectory, which was perceived as more difficult to travel than in the other conditions. This behaviour favoured a correct positioning of all the user's joints, reducing the kinematic error also at the shoulder and elbow levels.
The participants obtained significantly better performance with the use of our application. All the results come from tests conducted with healthy subjects, possessing normal abilities in executing physiological motor tasks and who therefore could be considered as not really in need of feedback support. We could consequently expect our system to produce a greater improvement with motor-impaired subjects with adequate cognitive functions. With them, we can also expect greater differences in the end-effector tracking performance between the three CIs.

Concerning the qualitative results coming from the healthy participants' questionnaires, we can say that the scores obtained are encouraging. The \textit{Comfort of the device} was ranked 4.04 out of 5, confirming the idea that AR headsets are an appreciated solution \citep{Jordan2011AugmentedStroke}. The guiding interface was highly positively evaluated (4.57/5) for its efficacy. Having an effective user interface is fundamental to ensure that the system is manageable also by subjects suffering from neurological diseases and, therefore, not necessarily practical when dealing with advanced technologies. One of the participants was colourblind and confirmed that, thanks to the presence of symbols on the buttons and the different sizes of the spheres composing the trajectories, the guiding interface was intuitive.
The category that obtained the lowest score is \textit{Reliability and stability of the hand-tracking system}: unluckily, the HoloLens hand-tracking algorithm suffers from some limitations when the hand is either too far or too close to the visor. Overall, the positive feedback from the healthy participants encourages potential future testing with patients.

The feedback coming from the clinical personnel confirmed the positive conclusions. The SUS score (68/100) ranked our system as "OK". Even if this value is slightly lower than the boundary for classification as "GOOD", we need to consider it in light of the SUS scores attributed by participants of other studies to applications developed with HoloLens. In \citep{Escalada-Hernandez2023UsabilitySkills}, some medical students rated a surgical training application in AR with a mean SUS score of 73.15. Hsieh \textit{et al.} \citep{Hsieh2020UsabilityAdults} developed an application for fall-risk assessments with older adults, and the participants of their study gave the system an average score of 71.9. Considering the standard SUS questions presented in \citep{Moosburner2019RealSurgery}, the Usability score gained by a surgical HoloLens-based application was 70. Our framework seems to be therefore in line with the average score of similar applications. It's important to note that these applications have different medical objectives and, so far, no other system has been developed for the same purpose as our framework.
Moreover, given the limited time the therapists had for training with our system and its present condition as a prototype, we believe there are high chances of improvement in terms of usability.
The results of the TAM are encouraging. The clinicians' Willingness To Use the system and the Perceived Usefulness both scored over 4 out of 5 points. PEOU scored 3.5 out of 5, which aligns with the responses from the SUS questionnaire. In fact, as we can notice from Fig. \ref{fig:sus}, the questions that lowered the final score are those related to the easiness and readiness of use ((3): \textit{The system is easy to use}, (4): \textit{I would need help to learn how to use the system}, (7): \textit{Most people would learn how to use the system easily}). Even though the system may not be seen as completely easy or intuitive to use, this does not significantly impact the intention to use it for rehabilitation sessions (PEOU-WTU correlation coefficient = 0.42). The low correlation between PEOU and WTU aligns with our hypothesis that the perceived complexity of the system can be addressed with additional training and is not currently viewed as a barrier to using the application. The multiple regression model confirms the prevalence of the effect of PU on WTU.
Both \textit{Importance of Tridimensionality} and \textit{Importance of Graphical aspects} obtained a final score of over 4 out of 5 points, confirming that graphics and details refinement are considered relevant to enhance users' engagement and compliance with the therapy. Both from the correlation analysis and the regression model we can see how Tridimensionality is considered as slightly more relevant for the utility of the system than the Graphical aspects.
 
We can notice how the Attitude Towards Technology of the participants has a positive and significant correlation with the scores of the $WTU$: therapists who are more confident with technologies tend to learn faster how to use the headset and are more interested in using it. Given also that it is demonstrated that good levels of digital literacy can enhance the benefits of the application of rehabilitation technologies \citep{Stark2023}, we can hypothesize that, in a future study on patients, participants with a good attitude towards technologies will experience greater improvements with our system. 
No disturbances related to motion sickness were reported, confirming that AR headsets could also be a promising solution to be tested with patients. In this sense, an interesting future step for this work could be testing the application with patients suffering from neurological disorders, to verify if the results we gathered with healthy subjects are confirmed. During our test, we interviewed the therapists regarding the types of patients who could benefit from using our system. They all recommended it for neurological patients who can engage with the therapist behaviourally and possess cognitive and attention levels that enable them to comprehend the required tasks. They should have already regained the motor abilities needed to execute functional gestures and just have to improve the accuracy of their movements.
\\
\\
We can then conclude, based on both the kinematic analysis and the therapists' evaluations, that the system we developed to provide three-dimensional feedback on the execution of therapeutic exercises has the potential to effectively support patients and therapists. The kinematic improvements obtained by our healthy participants while using the HoloLens feedback demonstrate its efficacy in enhancing tracking accuracy and, therefore, in potentially supporting rehabilitation outcomes in the long term for subjects with neurological disorders. The framework was appreciated both by healthy participants and clinicians. The obtained SUS score indicates that the system's usability is in line with other prototypes in the field of AR for healthcare and that, especially considering the high levels of perceived utility and intention to use it expressed by therapists, there is space for improvement and further exploration with patients.

\backmatter

%\bmhead{Supplementary information}

\bmhead{Acknowledgments}
We thank the staff from the Casa di Cura Igea S.p.A., for their invaluable participation in the study, and all colleagues and friends who volunteered and offered their time and interest.

\section*{Declarations}

%Some journals require declarations to be submitted in a standardised format. Please check the Instructions for Authors of the journal to which you are submitting to see if you need to complete this section. If yes, your manuscript must contain the following sections under the heading `Declarations':

\textbf{Funding}: not applicable. \\
\textbf{Conflict of interest/Competing interests}: FB, AP and MG hold shares in AGADE srl, Milan, Italy. The remaining author declares that the research
was conducted in the absence of any commercial or financial
relationships that could be construed as a potential conflict of
interest. \\
\textbf{Ethics approval and Consent to participate}: All the participants in the experiments signed an informed declaration of consent and the tests were approved by the Ethical Committee of Politecnico di Milano (approval n. 8/2022 - 16/02/2022). \\
\textbf{Consent for publication}: The questionnaires were developed in compliance with the Italian personal data protection law and the data were used exclusively for the purposes of study and scientific research and processed pursuant to Art. 13 of EU Regulation 2016/679 (General Data Protection Regulation). By responding to the questionnaire, the participants were informed and accepted that their answers were recorded and analyzed in aggregate form.\\
\textbf{Availability of data and materials and Code availability}: The datasets and codes used and/or analysed during the current study are available from the corresponding author on reasonable request. \\
\textbf{Authors' contributions}: BL and MG conceived the idea and carried out the experiments, BL developed the application, analyzed the data and wrote the manuscript, MG verified the analytical methods, PT and AS organised the experimental sessions with the clinicians and contributed to the final version of the manuscript, FB and AP supervised the work and contributed to the final version of the paper.

%%===================================================%%
%% For presentation purposes, we have included        %%
\bigskip

\begin{appendices}

\section{Questionnaire for the healthy participants}\label{AppA}

\textbf{Clarity of the calibration}
\begin{itemize}
    \item The process explanation is clear and understandable;
    \item The use of the index to locate the marker is not intuitive;
    \item The index tracking is stable;
    \item The button to fixate the marker is difficult to use.
\end{itemize}
\textbf{Usability of the trajectory manipulation mechanism}
\begin{itemize}
    \item The position of the trajectories is clearly identifiable and easy to reach;
    \item The desired trajectory is easy to select;
    \item The positioning mechanism is disorienting;
    \item The selection of the CIs is easy;
    \item The hand tracking to position the hand is stable.
\end{itemize}
\textbf{Effectiveness of the guiding interface}
\begin{itemize}
    \item It is easy to understand where the trajectory is and how to travel it;
    \item The colour code and the spheres help in understanding how to navigate the trajectory;
    \item The real trajectory is clearly visible;
    \item Colours and shadows help perceive the tridimensionality and depth of the objects.
\end{itemize}
\textbf{Reliability and stability of the hand-tracking system}
\begin{itemize}
    \item The recognition of the segments of the hand is always correct;
    \item I never lost track of the hand and had to change my position to have the system recognise it again;
    \item It is easy to interact with the buttons.
\end{itemize}
\textbf{Comfort of the device}
\begin{itemize}
    \item The headset is easy to use;
    \item I did not have difficulties in moving due to the headset encumbrance;
    \item Holograms colours were easy to recognise;
    \item The weight of the device is not too high to bother me;
    \item I had difficulties in seeing the holograms;
    \item I experienced nausea at least once during the use of the headset.
\end{itemize}

\section{TAM questionnaire for the clinical staff}\label{secA2}
\textbf{Willingness to use}
\begin{itemize}
    \item I think I would like to use this system for my patients;
    \item I believe I would likely use the system during my therapy sessions as a feedback system for my patients;
     \item I think I do not want to include the system in my therapy sessions;
     \item I do not think I am interested in using the headset.    
\end{itemize}
\textbf{Perceived Ease of Use}
\begin{itemize}
    \item I find the system difficult to use;
    \item I believe it is easy for the patient to learn how to use the system;
     \item I believe it is easy for me to learn how to use the system;
     \item Interacting with the system is often frustrating;
     \item I find it easy to manage the system.
\end{itemize}
\textbf{Perceived Usefulness}
\begin{itemize}
    \item I think the system is a valid instrument to support rehabilitation;
    \item I believe the system can help patients in the execution of exercises;
     \item I do not think that the feedback system is useful for the patient or helps them accurately exercise;
     \item I believe the system does not support any aspect of my rehabilitation work.    
\end{itemize}
\textbf{Tridimensionality}
\begin{itemize}
    \item I think the 3D feedback system is a useful addition with respect to other feedback systems;
    \item I do not think that tridimensional feedback improves the positive effects of therapy if compared to other simpler systems (just target points, vocal instructions...)
\end{itemize}
\textbf{Independence of the therapy}
\begin{itemize}
    \item I think that the autonomous reproduction of movements I taught has beneficial effects on the therapy;
    \item I think the system allows the patients to execute specific movements I want them to do, without my guidance.
\end{itemize}
\textbf{Graphics}
\begin{itemize}
    \item I think the patient benefits from the color-coded feedback;
    \item I appreciate the colour scale for its immediateness;
    \item I believe that the different dimensions of the trajectories are useful to improve the efficacy of the system;
    \item I do not think the colour scale helps me understand how well the patient is performing the task.
\end{itemize}
\end{appendices}

%%===========================================================================================%%
\bibliography{sn-bibliography}

\end{document}